\pdfoutput=1
\documentclass[10pt,american,a4paper,twocolumn]{article}
\usepackage{amsfonts,amsmath,amssymb}
\usepackage{braket}
\usepackage[font={footnotesize}]{caption}
\usepackage[T1]{fontenc}
\usepackage[margin=2cm]{geometry}
\usepackage{graphicx}
\usepackage{setspace}
\usepackage{times}
\usepackage{xspace}
\usepackage{hyperref}
\usepackage[square,super,comma,sort&compress]{natbib}

\makeatletter
\renewcommand\@makefntext[1]{\leftskip=2em\hskip-2em\@makefnmark#1}
\makeatother

\title{Separating \emph{Para} and \emph{Ortho} Water$^\ast$$^\ast$}
\author{%
   Daniel A. Horke, Yuan-Pin Chang, Karol D\l ugo\l\k{e}cki, and Jochen K\"upper$^{\ast}$\\}
\date{}
\begin{document}
\baselineskip24pt
\maketitle
\let\thefootnote\relax\footnotetext{[$^\ast$]%
   \hspace*{1.5ex}Dr. D.A. Horke, Dr. Y.-P. Chang, K. D\l ugo\l\k{e}cki, Prof. Dr. J. K\"upper\\
   Center for Free-Electron Laser Science, DESY\\
   Notkestrasse 85, 22607 Hamburg (Germany)\\
   E-mail:  jochen.kuepper@cfel.de\\
   http://desy.cfel.de/cid/cmi/\\ \\
   Prof. Dr. J. K\"upper\\
   The Hamburg Center for Ultrafast Imaging\\
   Luruper Chaussee 149, 22761 Hamburg (Germany)\\
   Department of Physics, University of Hamburg\\
   Luruper Chaussee 149, 22761 Hamburg (Germany)\\}
\let\thefootnote\relax\footnotetext{[$^\ast$$^\ast$]%
   We acknowledge experimental contributions by Antoine Moulet in the early stage of the experiment.
   In addition to DESY, this work has been supported by the excellence cluster ``The Hamburg Center
   for Ultrafast Imaging -- Structure, Dynamics and Control of Matter at the Atomic Scale'' of the
   Deutsche Forschungsgemeinschaft.}%
\singlespacing%
Water exists as two nuclear-spin isomers, \emph{para} and \emph{ortho}, determined by the overall
spin of its two hydrogen nuclei. For isolated water molecules the conversion between these isomers
is forbidden and they act as different molecular species. Yet, these species are not readily
separated and no pure \emph{para} sample has been produced. Accordingly, little is known about their
specific physical and chemical properties, conversion mechanisms, or interactions. Here, we
demonstrate the production of isolated samples of both spin isomers in pure beams of \emph{para} and
\emph{ortho} water in their respective absolute ground state. These single-quantum-state samples are
ideal targets for unraveling spin-conversion mechanisms, for precision spectroscopy and
fundamental-symmetry-breaking studies, and for spin-enhanced applications, e.\,g., laboratory
astrophysics and \mbox{-chemistry} or hypersensitized NMR experiments.

Significant efforts have been undertaken to separate and study the nuclear-spin isomers of water,
motivated by their importance in a wide variety of scientific disciplines. This ranges from the
astronomical importance of the \emph{ortho}-\emph{para} ratio~\cite{Mumma:Science232:1523,
   Hogerheijde:Science334:338, Dishoek:ChemRev113:9043, Tielens:RevModPhys85:1021,
   Lis:JPCA117:9661}, to studies of nuclear-spin conversion~\cite{Curl:JCP46:3220,
   Chapovsky:ARPC50:315}, selection rules and reactive collisions~\cite{Quack:MP34:477,
   Oka:JMS228:635, Uy:PRL78:3844} or symmetry breaking~\cite{Mazotti:PRL86:1919}. Spin-enriched
samples furthermore would allow for hypersensitized NMR experiments \emph{via} polarization transfer
reactions~\cite{Bowers:PRL57:2645, Bouchard:Science319:442, Emondts:PRL112:077601}. However, unlike
other small polyatomic molecules exhibiting spin isomerism, such as fluoromethane or
ethylene~\cite{Krasnoperov:JETPLett39:143, Sun:Science310:1938}, which were spin-isomerically
enriched using the light induced drift technique~\cite{Chapovsky:ARPC50:315}, this has not been
achieved for water. Separation through selective adsorption on surfaces was reported~
\cite{Tikhonov:Science296:2363}, but these results remain controversial and could not be
reproduced~\cite{Limbach:ChemPhysChem7:551, Veber:JETP102:76, Buntkowsky:ZPhysChem222:1049,
   Cacciani:PRA85:012521}. Thus, studies of spin-conversion dynamics in water have been limited to
water embedded in rare gas matrices, with relative spin populations modified by the sample
temperature~\cite{Abouafmarguin:CPL447:232}. A recent study investigated nuclear-spin conversion in the
gas-phase and found no spin conversion for water monomers~\cite{MancaTanner:JPCA2013}.

Recently, the production of a single spin isomer of water in a magnetic-hexapole-focuser setup was
demonstrated~\cite{Kravchuk:Science331:319}. One of the magnetically active spin projections
($m=+1$) of ground-state \emph{ortho} water was magnetically focused into the interaction volume,
while all other spin-projection states were defocused or diverged unaffected by the field. The
purity of the produced \emph{ortho} beam was later evaluated as 93\,\%, with simulations suggesting
an upper limit for the achievable purity of 97\,\%~\cite{Kravchuk:Science331:319,
   Turgeon:PRA86:062710}.

Here, we experimentally demonstrate the production of pure samples of both, \emph{para} and
\emph{ortho} water, the latter further separated into its $M=0$ and $M=1$ angular momentum
projections, in the gas-phase. The produced single-quantum-states are ideally suited for further
experiments on nuclear-spin conversion under collision-free conditions, nuclear-spin-dependent
reactivity~\cite{Quack:MP34:477}, trapping of single spin-isomer samples in electromagnetic
traps~\cite{Meerakker:CR112:4828} or cold matrices~\cite{Turgeon:PRA86:062710}.

\begin{figure}
   \centering
   \includegraphics[scale=0.8]{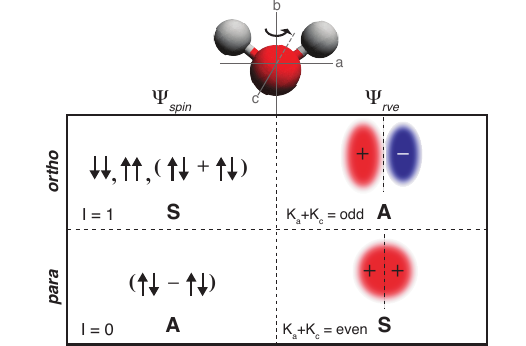}
   \caption{Water spin isomers. The necessity for an overall antisymmetric wavefunction requires
      that the symmetric ($S$) spin combinations $\Psi_\text{spin}$ of \emph{ortho} water combine
      with an antisymmetric ($A$) rovibronic eigenstate $\Psi_{rve}$, and \emph{vice versa} for
      \emph{para} water.}
   \label{fig:intro}
\end{figure}
Nuclear-spin isomers are different molecular species that arise from the indistinguishability of
identical protons, each of which can have its nuclear spin $(i=$1/2$)$ \emph{up} $(m_i=+$1/2$)$ or
\emph{down} $(m_i=-$1/2$)$. In the case of water the nuclear spins of the two equivalent protons can
be combined in four different ways, shown in \autoref{fig:intro}. These combinations are grouped
into one antisymmetric and three symmetric nuclear-spin wavefunctions, termed \emph{para} ($I=0$)
and \emph{ortho} ($I=1$), respectively. The symmetrization postulate (Pauli principle) requires an
overall antisymmetric wavefunction with respect to exchange of the two fermionic hydrogens. This
constrains the allowed combinations of rovibronic eigenstates ($\Psi_\text{rve}$) with spin
configurations ($\Psi_\text{spin}$), i.\,e., the product of the two corresponding symmetry species
must be antisymmetric regarding this exchange. Under the conditions of a cold molecular beam, all
molecules reside in the ground electronic and vibrational state, both of which are totally
symmetric. Therefore, the restrictions are on the rotational levels corresponding to each spin
isomer. Relative abundances are determined by the spin degeneracy for each symmetry.

\begin{figure}
   \centering
   \includegraphics[scale=0.8]{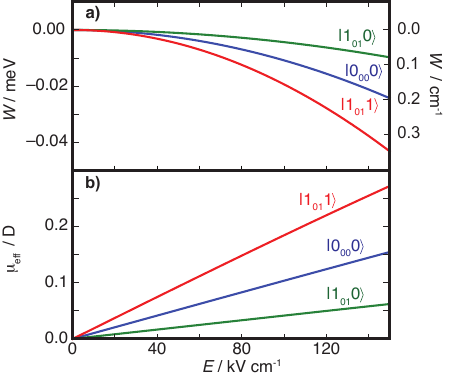}
   \caption{Calculated a) Stark-energy shifts $W$ and b) effective electric dipole moments
      $\mu_{\text{eff}}$ for the absolute ground states of \emph{para} and \emph{ortho} water as a
      function of electric-field strength $E$.}
   \label{fig:stark}
\end{figure}
To spatially separate the spin isomers we exploit the different rotational states occupied by
\emph{para} and \emph{ortho} water and the corresponding differences in their dc Stark effects. The
rotational quantum states of the asymmetric rotor, i.\,e., water, can be classified by
$J_{K_aK_c}M$, with the total angular momentum quantum number $J$, the projection labels $K_a$ and
$K_c$ onto the molecule-fixed $a$ and $c$ axes as defined in \autoref{fig:intro}, respectively, and
the projection quantum number $M$ onto the space-fixed $Z$ axis. In the molecule-fixed coordinate
system exchange of the proton spins corresponds to rotation of $\pi/2$ about the $b$ axis, which is
identical to a rotation of $\pi/2$ about $a$ followed by a rotation of $\pi/2$ about $c$. The
corresponding symmetry of the rotational wavefunction is the product of the parities,
$P=(-1)^{K_a+K_c}$. This leads to \emph{para} water requiring $K_a+K_c$ being even with an absolute
ground state of $\ket{0_{00}0}$. For \emph{ortho} $K_a+K_c$ is odd and the ground state is denoted
$\ket{1_{01}M}$. The responses of the two absolute ground states to a strong dc electric field are
displayed in \autoref{fig:stark}\,a, showing the non-degeneracy of the \emph{ortho} water $M$ states
in the presence of an electric field. The differences in the Stark effect lead to distinct effective
dipole moments $\mu_{\text{eff}}$, i.\,e., space-fixed dipole moments, plotted in
\autoref{fig:stark}\,b.

The force experienced by the molecules inside the electric deflection field $E$ is proportional to
$\mu_{\text{eff}}\cdot\nabla{E}$~\cite{Filsinger:JCP131:064309}. The ground states of both
\emph{para} and \emph{ortho} water are strong-field seeking (\autoref{fig:stark}), preventing their
separation using electric focusing techniques for weak-field-seeking
states~\cite{Meerakker:CR112:4828}. The general applicability of the electrostatic deflection
technique to water samples was demonstrated.~\cite{Scheffers:PhysZ40:1, Moro:PRA75:013415} Our
combination of the deflection by strong inhomogeneous electric fields with very cold molecular beams
allows for the separation of the nuclear-spin states. The quantum-state-resolved detection method
allows for their unambiguous assignment.
\begin{figure}
   \centering
   \includegraphics[scale=0.8]{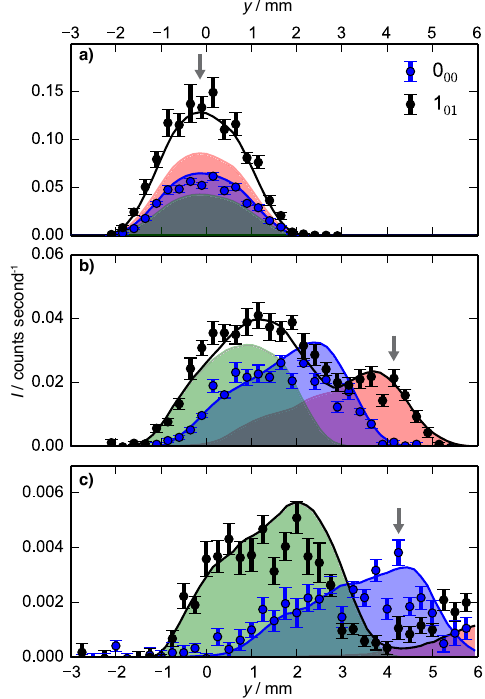}
   \caption{Measured (data markers) and simulated (solid lines) spatial profiles for water
      co-expanded in 40~bar of neon using deflection voltages of a) 0~kV and b) 15~kV. c) Expansion
      in 15~bar of argon using a deflection voltage of 15~kV. Green and red shading correspond to
      $M=0$ and $M=1$ levels of the $ \ket{1_{01}}$ state, respectively and blue shading to the $
      \ket{0_{00}}$. Gray arrows indicate the positions at which the spectra in Fig. 4 were
      collected.}
   \label{fig:spatial}
\end{figure}
In the current experiment (see supporting information for details) a supersonic molecular beam is
used to produce a cold water sample in the gas-phase with a rotational temperature of $8$~K,
corresponding to $>\!99$\,\% of \emph{para} and $>\!96$\,\% of \emph{ortho} molecules in their
absolute ground state, respectively. The molecular beam is then dispersed perpendicular to its
flight direction according to the effective-dipole-moment-to-mass ratio ($\mu_{\text{eff}}/m$) using
strong inhomogeneous electric fields~\cite{Filsinger:JCP131:064309,Filsinger:ACIE48:6900}. Water
molecules are quantum state selectively ionized \emph{via} (2+1) resonance-enhanced multiphoton
ionization (REMPI); a spectrum is shown in the supporting information.

Spatial profiles of the individual quantum states in the molecular beam are shown in
\autoref{fig:spatial}, with solid lines indicating corresponding trajectory simulations (see
supporting information for details). In the absence of a deflection field, \autoref{fig:spatial}\,a,
the \emph{para} and \emph{ortho} constituents of the beam are mixed and centered around the zero
position.

\begin{figure}
   \centering
   \includegraphics[scale=0.8]{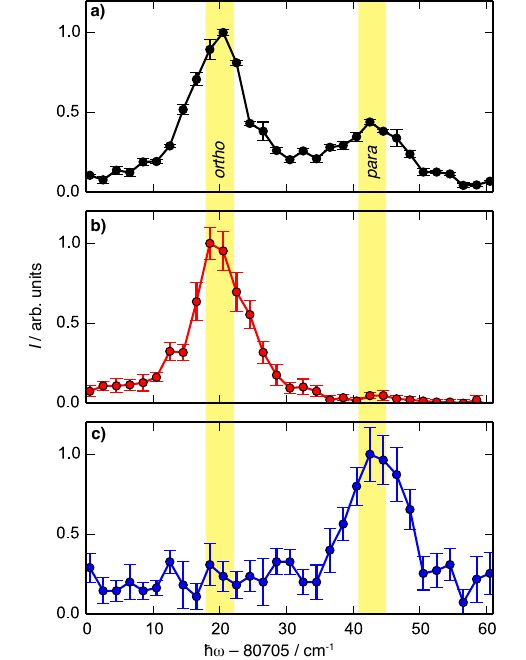}
   \caption{Rotationally resolved 2+1 REMPI spectra showing characteristic transitions for
      \emph{para} and \emph{ortho} water. a) Undeflected beam with thermal population of both spin
      states. b) Pure \emph{ortho} ($\ket{1_{01}1}$) sample created in the neon expansion at 15~kV
      deflection voltage. c) Pure \emph{para} ($\ket{0_{00}0}$) sample created in the argon
      expansion at 15~kV deflection voltage.}
   \label{fig:spectra}
\end{figure}

This is confirmed by the REMPI spectrum obtained at this position, \autoref{fig:spectra}\,a.
Analysis of the spectrum yields a $\emph{para}:\emph{ortho}$ ratio of approximately $1:3$,
consistent with a conservation of the nuclear-spin temperature through the supersonic expansion.
Application of an electric field deflects the beam in the upward direction. The spatial shift
depends on the effective dipole moment, and a clear separation of \emph{para} and \emph{ortho} water
is observed. At large deflection fields the spatial profile of \emph{ortho} water bifurcates,
corresponding to the splitting of the $M=0$ and $M=1$ components. This is indicated by the green and
red shading in \autoref{fig:spatial}\,b and c. Consistent with the calculated Stark curves and
effective dipole moments, \autoref{fig:stark}\,a and b, the $\ket{1_{01}1}$ state experiences the
largest deflection and the $\ket{1_{01}0}$ state the least deflection, with the $\ket{0_{00}0}$
\emph{para} state in between. For water co-expanded in neon, application of 15~kV to the deflector
creates a region of space, $y>4$~mm, where only the $\ket{1_{01}1}$ state is present and a pure
\emph{ortho} sample is obtained. This is confirmed by the REMPI spectrum shown in
\autoref{fig:spectra}\,b. The measured purity of the \emph{ortho} beam at this position is 97\,\%,
primarily limited by background water in the vacuum chamber. Simulations suggest an achievable
purity in excess of 99\% with the present setup; see supporting information for details.

Further increasing the deflection through the use of a slower molecular beam, seeded in argon, leads
to a nearly complete depletion of the $\ket{1_{01}1}$ \emph{ortho} state and the creation of an
enriched \emph{para} $\ket{0_{00}0}$ sample, as shown in \autoref{fig:spatial}\,c and confirmed by
the REMPI spectrum, \autoref{fig:spectra}\,c. A purity of 74\,\% for \emph{para} water is measured,
which simulations indicate would be increased to $>87$\,\% if background water was more efficiently
suppressed. Using a setup with two subsequent deflection stages a purity $>96$\,\% could be
obtained. The slower beam, furthermore, allows the creation of a pure \emph{ortho} sample in the
$M=0$ angular momentum projection at a position $-1<y<0$~mm. This position is depleted of all other
quantum states in the original beam and we obtain a $>$99\,\% pure $\ket{1_{01}0}$ sample.

The produced pure molecular beams have densities on the order of $10^8$~cm$^{-3}$ and
$10^7$~cm$^{-3}$ for neon and argon expansions, respectively. The latter is limited by the longer
gas-pulse duration and possibly by the reverse seeding effect resulting from Argon being heavier
than water. Combined with the strong deflection experienced in the slower beam, this leads to
molecules colliding with the skimmers or electrodes and not reaching the interaction region anymore.
These densities are sufficient for precision spectroscopy or laboratory scattering
experiments~\cite{Curl:JCP46:3220, Chapovsky:ARPC50:315}.

The current experiment, at 20~Hz, allows the production of $\sim\!10^{13}$ nuclear-spin-selected
molecules, or 1~picoliter, per day. Significantly larger quantities could be produced using
higher-repetition-rate~\cite{Trippel:MP111:1738} or continuous molecular beams. The current
quantities could be sufficient for the production of nuclear-spin-pure surface layers, for instance,
highly polarizable layers of \emph{ortho} water, if very low spin-relaxation rates can be
maintained. The production of pure \emph{para} water might open up possibilities for hypersensitized
NMR experiments through polarization transfer \emph{via} water addition, comparable to
\emph{para}-hydrogen induced polarization transfer (PHIP)~\cite{Bowers:PRL57:2645,
   Bouchard:Science319:442}, but with wider chemical applicability.

The presented technique for quantum-state separation is generally applicable to polar neutral
molecules and allows for the production of single-quantum-state samples, i.\,e., for the separation
of nuclear-spin isomers.


\footnotesize
\vspace{-3em}
\bibliography{string,cmi}
\bibliographystyle{Angewandte}
\end{document}